\documentclass[twocolumn,prd,floatfix,nofootinbib]{revtex4}

\usepackage{mathrsfs}
\usepackage{latexsym}
\usepackage{amsmath}
\usepackage{graphicx}
\usepackage{subfigure}
\usepackage{dcolumn}
\usepackage{bm}
\usepackage{amssymb}
\usepackage{latexsym}
\usepackage{ulem}
\usepackage{color}
\usepackage[colorlinks,linkcolor=magenta,anchorcolor=cyan,citecolor=blue]{hyperref}

\def\be{\begin{equation}}
\def\ee{\end{equation}}
\def\ba{\begin{eqnarray}}
\def\ea{\end{eqnarray}}
\def\red{\color{red}}

\def\nn{\nonumber}

\begin{document}

\begin{flushleft}
{\footnotesize
Imperial/TP/2019/JZ/01\\
USTC-ICTS-19-06
}
\end{flushleft}

\title{On echo intervals in gravitational wave echo analysis}

\author{Yu-Tong Wang$^{1}$\footnote{wangyutong@ucas.ac.cn}}
\author{Jun Zhang$^{2}$$\footnote{jun.zhang@imperial.ac.uk}$}
\author{Shuang-Yong Zhou$^{3}$\footnote{zhoushy@ustc.edu.cn}}
\author{Yun-Song Piao$^{1,4}$\footnote{yspiao@ucas.ac.cn}}

\affiliation{$^1$ School of Physics, University of Chinese Academy
of Sciences, Beijing 100049, China}

\affiliation{$^2$Theoretical Physics, Blackett Laboratory, Imperial College, London, SW7 2AZ, United Kingdom}

\affiliation{$^3$  Interdisciplinary Center for Theoretical Study,
University of Science and Technology of China, Hefei, Anhui 230026, China}

\affiliation{$^4$ Institute of Theoretical Physics, Chinese
Academy of Sciences, P.O. Box 2735, Beijing 100190, China}

\begin{abstract}
Gravitational wave echoes, if they exist, could encode important information of new physics from the strong gravity regime. Current echo searches usually assume constant interval echoes (CIEs) a priori, although unequal interval echoes (UIEs) are also possible. Despite of its simplicity, the using of CIE templates need to be properly justified, especially given the high sensitivity of future gravitational wave detectors. In this paper, we assess the necessity of UIE templates in echo searches. By reconstructing injected UIE signals with both CIE and UIE templates, we show that the CIE template may significantly misinterpret the echo signals if the variation of the interval is greater than the statistical errors of the interval, which is further confirmed by a Bayesian analysis on model stelection. We also forecast the constraints on the echo intervals given by future GW detectors such as Advanced LIGO and Einstein Telescope.
\end{abstract}

\maketitle

\section{Introduction}

The Laser Interferometer Gravitational Wave Observatory (LIGO) and Virgo experiments have already successfully detected a dozen of compact binary coalescence events \cite{Abbott:2016blz,TheLIGOScientific:2017qsa,LIGOScientific:2018mvr}, which has opened up a new window into the strong gravity regime. As the most extremely compact objects characterized by the existence of event horizons, black holes are some of the most intriguing objects to be tested with gravitational wave (GW) observations. Although the events observed so far are compatible with black holes predicted in general relativity (GR), it would be still premature to declare the existence of the black hole event horizon \cite{Cardoso:2017cqb}.

In particular, exotic compact objects (ECOs) \cite{Cardoso:2017cqb} without horizons \cite{Visser:2009pw, Cardoso:2016oxy,  Cardoso:2017njb, Cardoso:2017cfl, Holdom:2016nek, Zhang:2017jze,Maggio:2017ivp,Raposo:2018xkf,Ghersi:2019trn,Maggio:2018ivz,Maselli:2018fay} can still be potential candidates that are responsible for the GW events. Examples of ECOs include gravastars \cite{Mazur:2004fk,Visser:2003ge} and boson stars \cite{Liebling:2012fv, Brito:2017wnc, Palenzuela:2017kcg}, as well as alternatives to GR black holes like fuzzballs \cite{Mathur:2005zp}, which are motivated by quantum gravity considerations and in attempts to address the black hole information paradox \cite{Almheiri:2012rt, Polchinski:2016hrw, Maldacena:2013xja, Lunin:2001jy, Lunin:2002qf}. Such ECOs usually have surfaces that are assumed to be slightly larger the would-be black hole horizon. In this case, GWs scattered on a horizonless ECO will not be entirely absorbed as they are scattered on a GR black hole. Instead, some of the GWs will be reflected by the ECO's surface and form GW echoes. Despite its exotic nature, ECOs are still viable candidates to explain the LIGO GW events. As shown in \cite{Cardoso:2016rao}, the initial ringdown signals, such as those observed by LIGO so far, only reflect the geometry near the photon sphere. In other words, the initial ringdown signal from an ECO can be very similar to that of a GR black hole, if the surface of the ECO is deep inside the photon sphere. GWs reflected by the ECO's surface, i.e., the echoes, only show up in the GW signals at a later stage. Thus if they can be detected, GW echoes will be evidence of ECOs and a good probe to the physics near the ECO's surface.

Physically, echoes in the late-time ringdown signal are caused by repeated reflections between the ECO's surface and the potential barrier at the photo sphere. The time interval between two successive echoes marks the scale of the new physics, and thus is an important quantity to consider in search strategies \cite{Correia:2018apm}. The waveform templates used in current echo searches usually assume constant interval echoes (CIEs). However, given the various possibilities and unknown nature of the structure of the ECO surface and the ringdown dynamics, unequal interval echoes (UIEs) may also appear. For example, it has been shown in Ref \cite{Wang:2018mlp} that, if the post-merger object is a wormhole that is slowly pinching off and afterwards collapses into a black hole, the late-time ringdown waveform will exhibit a train of echoes with increasing-intervals. Moreover, a characteristic unequal time interval may also indicate an exotic origin of the object from the inflation period in the early universe \cite{Wang:2018cum}. Therefore, the use of the CIE template may not be always justified.

In this paper, we investigate the necessity of using the UIE template in echo searches. We consider a fiducial scenario with UIEs, and simulate the data by injecting an UIE signal to the noise generated by the forecast noise curve for Advanced LIGO at design sensitivity. We assume that the echo interval increase with a constant ratio $r$. By performing Markov Chain Monte Carlo (MCMC) sampling, we first show that the CIE template fails to extract echo signals if the physical signals have unequal intervals with $r > \epsilon_{\Delta t_{echo}}$, where $\epsilon_{\Delta t_{echo}}$ is the relative error of parameter inference on the echo interval. To further understand the result, we perform the Bayesian model selection analysis. We investigate how the Bayes factor varies with different parameters, such as the amplitude of GW echoes or increment of time interval. A same approach is utilized by \cite{Veitch:2014wba, Westerweck:2017hus, Tsang:2018uie, Nielsen:2018lkf, Lo:2018sep} with PyCBC Inference for Bayesian model selection. However, these works focused on echoes with constant time interval. We also forecast the detectability of further GW detectors on GW echoes. The rest of the paper is organised as follows. In Sec.~\ref{Sec:Waveform templates}, we first introduce waveform templates that describe the CIE and UIE models. In Sec.~\ref{Sec:Parameter Inference}, we show the MCMC example, and perform the analysis on model selection. In Sec.~\ref{Sec:Fisher_matrix}, we use the Fisher information matrix to estimate the relative error on some key parameters in the CIE and UIE template and to show the results. In Sec.~\ref{Sec:Discussions}, we discuss the implication of our results on echo searches as well as on understanding ECOs.

\section{Waveform templates}\label{Sec:Waveform templates}

We first introduce the waveform templates that will be used to in the following analysis. Many efforts have been made towards constructing analytical templates that characterize the late time gravitational waveform from perturbed exotic objects \cite{Abedi:2016hgu,Maselli:2017tfq,Wang:2018gin,Testa:2018bzd,Ashton:2016xff,Abedi:2017isz, Burgess:2018pmm,Konoplya:2018yrp,Oshita:2018fqu}. As the main focus of this paper is on the echo intervals, it is sufficient to use a simple phenomenological waveform template:
\ba
\Psi(t) & =& \Psi^{BH}(t)+\Psi^{echo}(t)\nn\\
&=& Ae^{- t/{\tau}}\cos(2\pi {f}t+{\phi})
\nonumber \\
&&+ \sum_{n=1}^{{\tilde N}_{echo}}(-1)^{n}{
A}_{n}e^{-\frac{x^2_{n}}{2 \beta^2_{n}}}\cos(2\pi {f}_n
x_n). \label{template}
\ea
Here $\Psi^{BH}(t)$ is the post-merger black hole-like waveform with amplitude $A$, damping time $\tau$, reference phase $\phi$, and central frequency $f$.  $\Psi^{echo}(t)$ is a fiducial echo waveform, where $\tilde N_{echo}$ is the number of echoes after the main merger signal. For simplicity, we choose $\beta_n\sim\beta$, so that each echo has a Gaussian profile with a constant width, and ${A}_n\sim {{\cal A}A}/(3+n)$ , and ${\cal A}$ is the ratio of the amplitude of the first echo relative to $A$, up to an ${\cal O}(1)$ factor. We also fix the value of $f_{n}$ as central frequency $f$. The form of $x_n$ is different for the CIE and UIE templates.

\vskip 5pt
\noindent$\bullet$  \noindent {\bf CIE template.} If the time interval between the main merger signal and the first echo is $t_{echo}$, we have
 \be
 x_n = t-t_{echo}-n\Delta t_{echo}.
 \ee
The parameter $\Delta t_{echo}$ specifies the constant time interval between two successive echoes, encapsulating the compactness of the exotic compact object that is being probed. Physically, $\Delta t_{echo}$ directly relates to the distance between the photosphere potential barrier and the (reflection) surface of the ECO.

\begin{figure}[tbp]
\includegraphics[width=0.45\textwidth]{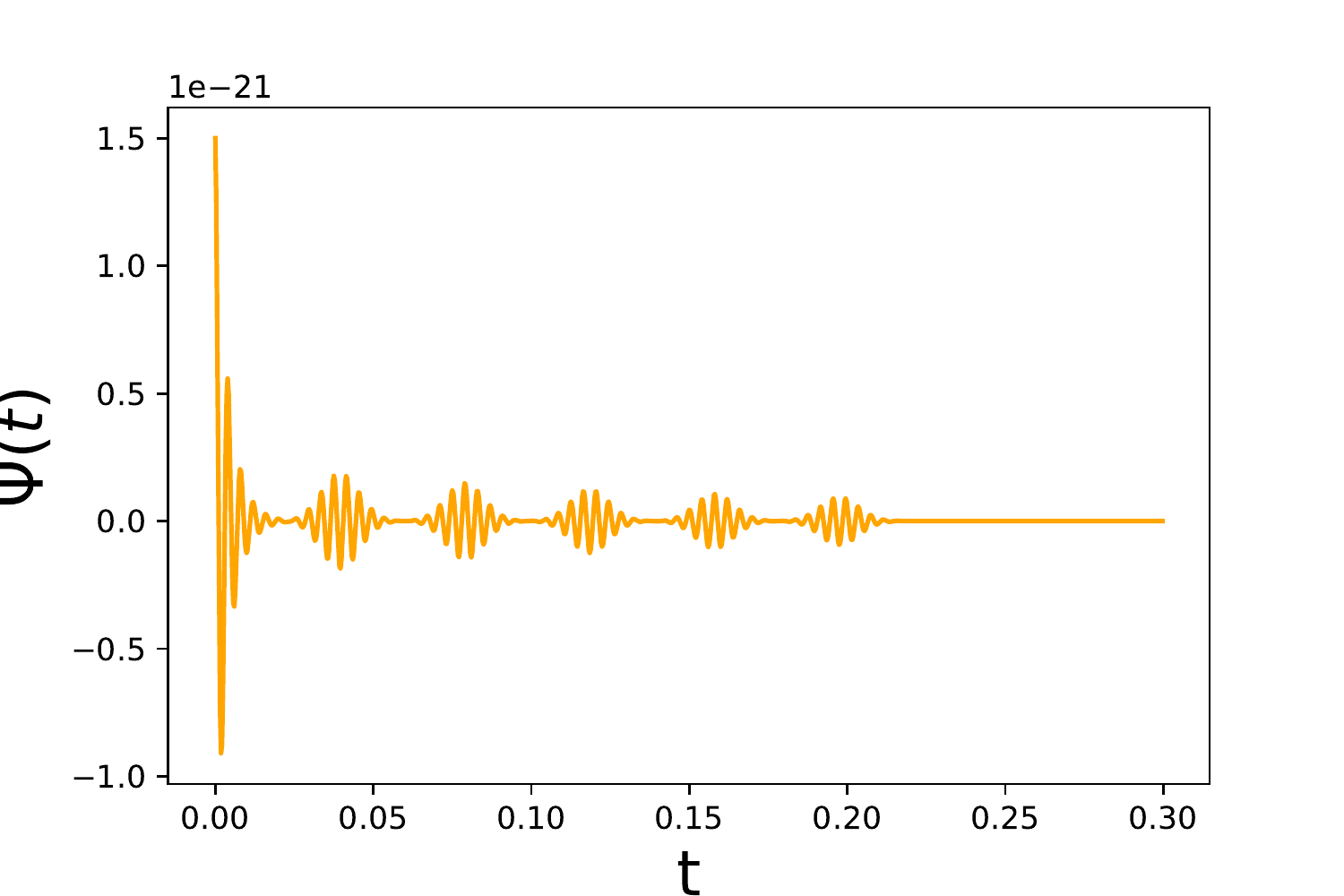}
\includegraphics[width=0.45\textwidth]{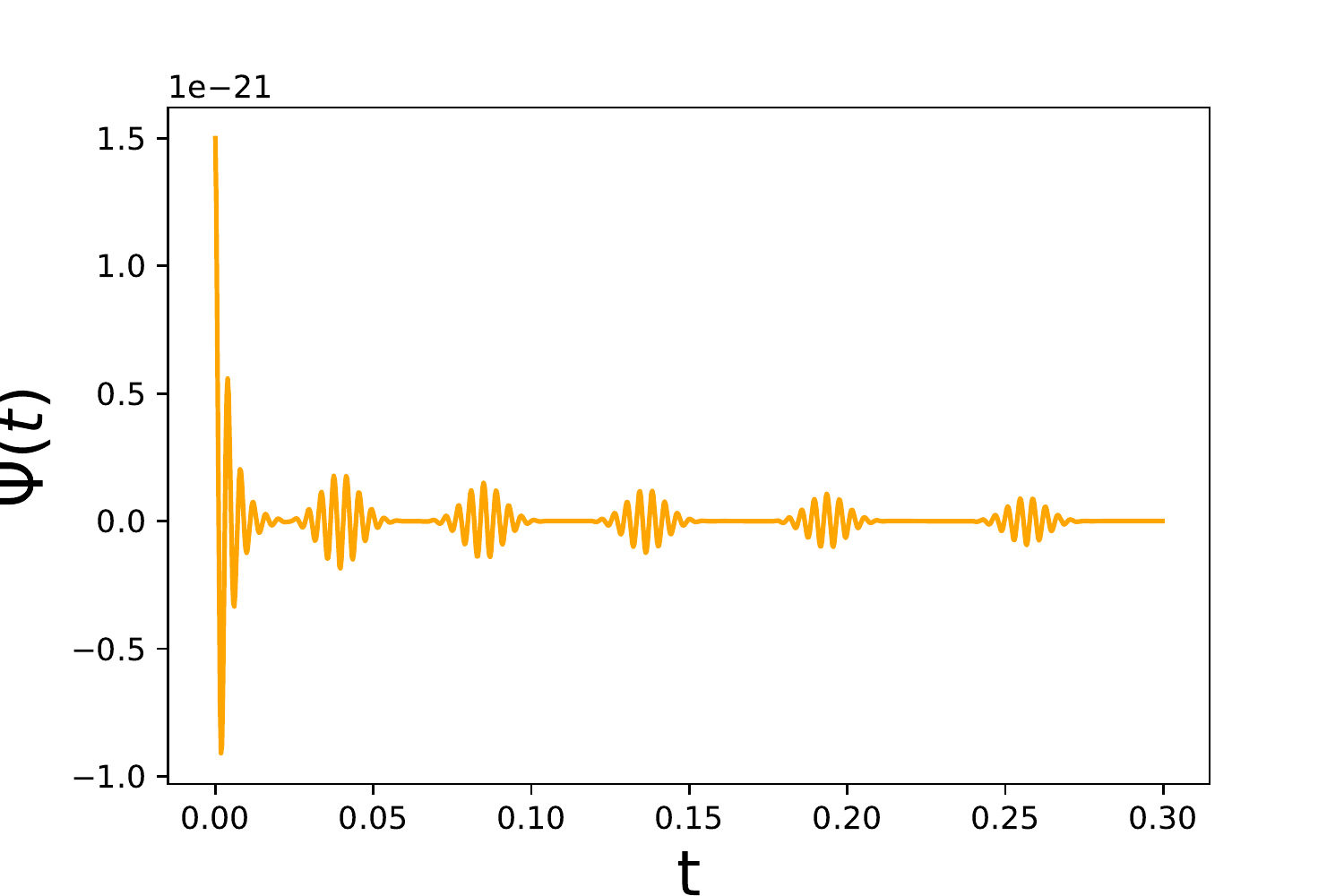}
\caption{GW echoes waveform templates. The horizontal axis is observation time and the vertical axis is the waveform amplitude strain. The top panel shows the CIE waveform and the bottom panel shows the UIE waveform. We choose  $\tau = 4\times10^{-3}$s,  $f = 250{\rm Hz}, \phi = 0,  A = 1.5\times10^{-21}, {\cal A} = 0.5, \beta = 0.006$s, $\Delta t_{echo}= 0.0295$s, $r=0.15$, and $t_{echo} = 0.0295$s.}
\label{fig01}
\end{figure}

\vskip 5pt
\noindent$\bullet$  \noindent {\bf  UIE template.} {\it A priori} the variation of time intervals between echoes can be very generic. For the illustrative purposes, however, we will focus on a simple case where the echo time interval increases monotonically and the increment is proportional to $\Delta t_{echo}$: $\delta t=r \Delta t_{echo}$ with $r$ being constant. In this case, we have
\be
\label{eq:xnr}
 x_n = t-t_{echo}-n\Delta t_{echo} - \frac{n(n+1)}{2}r\Delta t_{echo}.
 \ee
We expect that the extra $r\Delta t_{echo}$ term could arise from finer structures or dynamical features of the ECO surface, and can be a probe to some specific astrophysical and/or cosmology scenarios. Fig.~\ref{fig01} shows two examples of the above waveform templates.

\section{Parameter Inference and model selection}\label{Sec:Parameter Inference}

In this section, we show that for more generic echo signals, using the CIE template may be too restrictive, and can fail to extract the physical signals from observed data. GW signals are usually extracted by matched filtering. Assuming a given detection $d(t)$ consists of noise $n(t)$ and a physical signal that is modelled by a template $h(t;\,\theta)$, the likelihood function is given by
\be
\label{likelihood}
p(d|\theta,{\cal H},I) = {\cal N}\exp\left[-\frac{1}{2}\left\langle d-h | d - h\right\rangle\right]
\ee
where ${\cal N}$ is a normalization constant. Moreover, inference on the parameters may also depend on ${\cal H}$, the hypothesis that we choose, and $I$, the knowledge known prior to the selection. In practice, the likelihood function is calculated in the frequency-domain, in which case the inner product is defined as
\ba
\langle a|b \rangle = 2 \int_{0}^{\infty} df \dfrac{\tilde{a}^{*}(f) \tilde{b}(f) + \tilde{a}(f) \tilde{b}^{*}(f)}{S_{n}(f)},
\ea
where $\tilde{a}$ and $\tilde{b}$ are the Fourier transform of $a$ and $b$, and $S_n\left(f\right)$ is the detector noise spectral density. The inner product is defined so that the probability of a noise realzation $n_0(t)$ is $p(n=n_0)\propto \exp\left[-\langle n_0|n_0 \rangle /2\right]$

\begin{figure*}[htbp]
\includegraphics[width=0.485\textwidth]{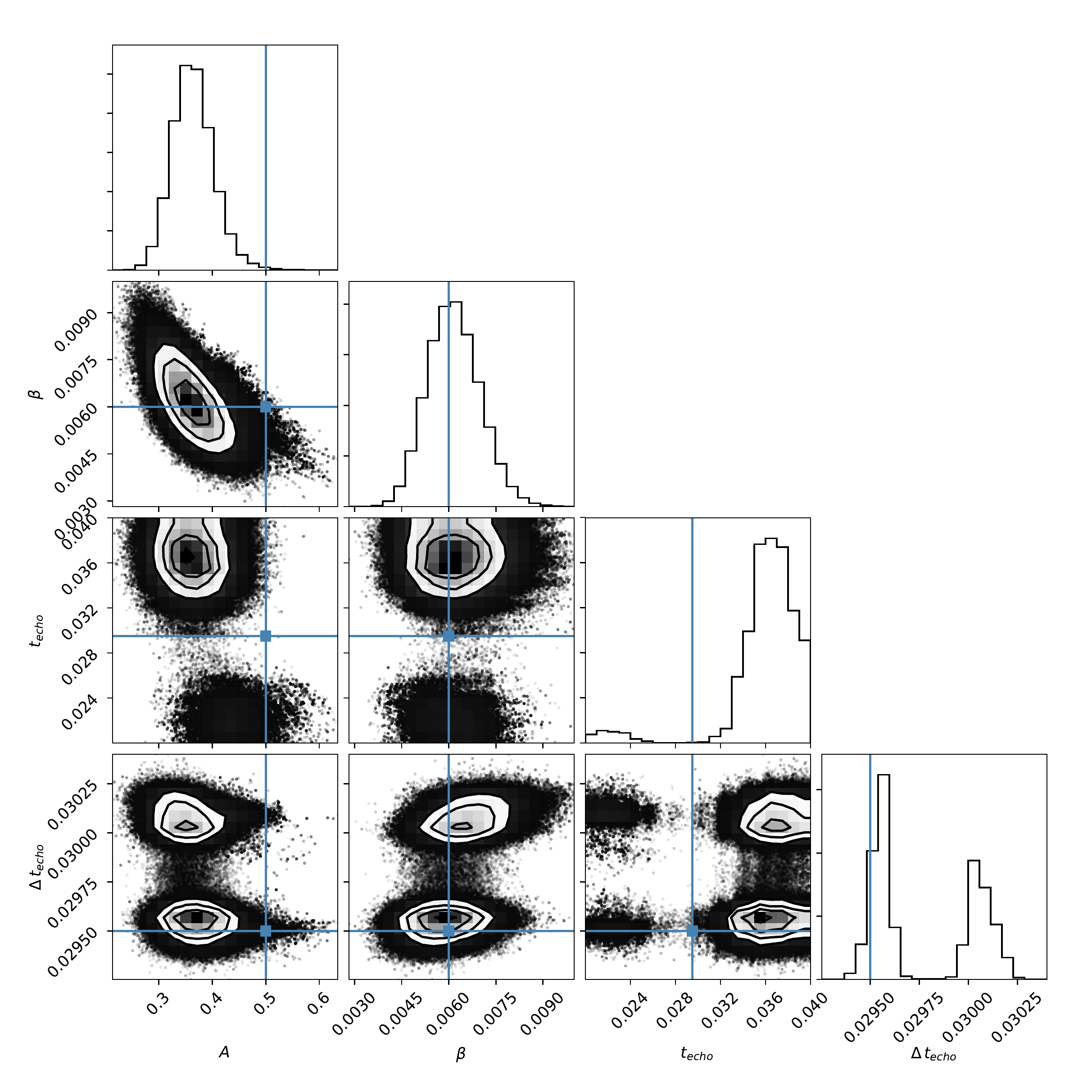}
\includegraphics[width=0.49\textwidth]{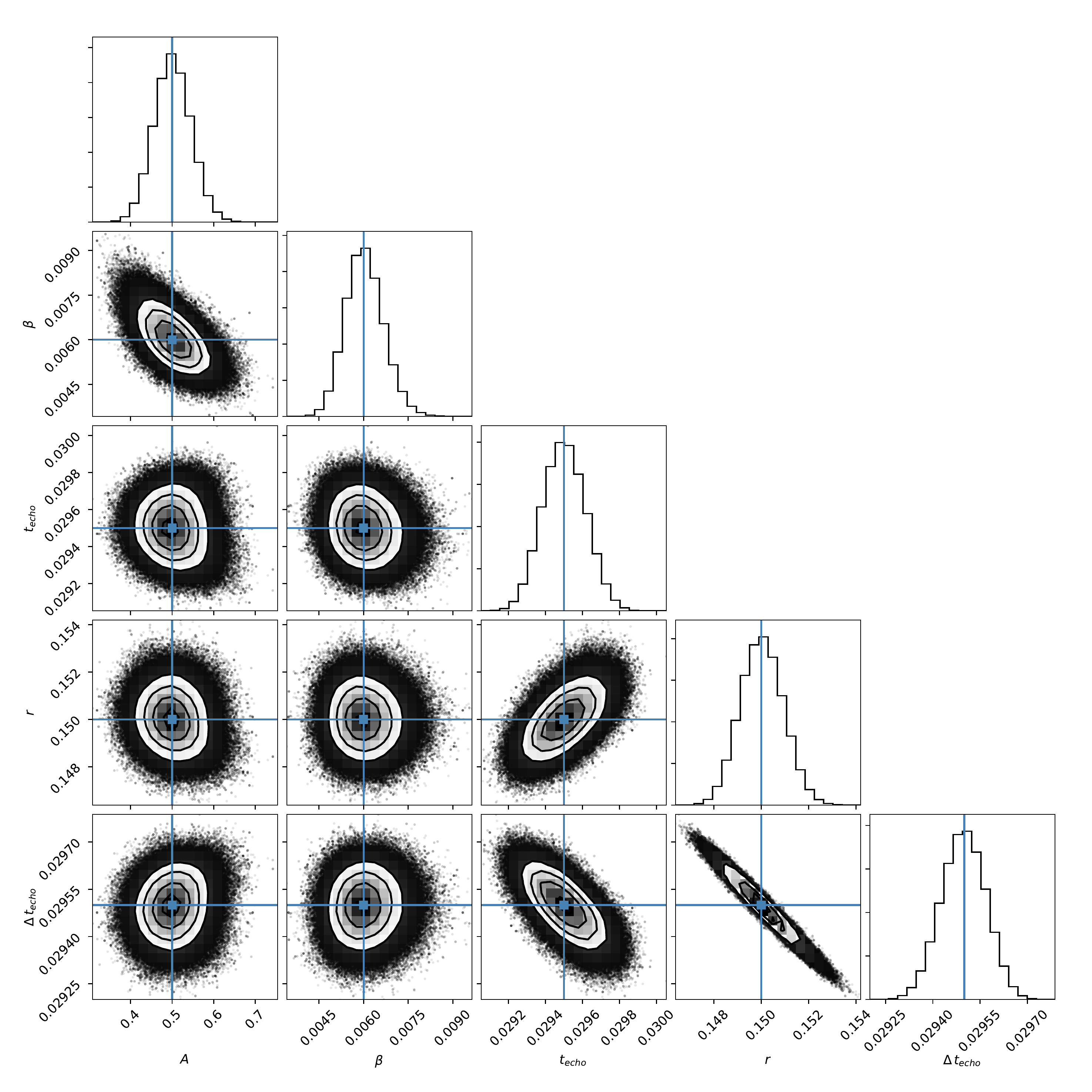}
\caption{Corner plots (left panel for the CIE model and right panel for the UIE model) of the posterior samples from the parameter estimation on simulated data as describe in the main text. The blue solid lines denote the injected values for the respective parameters. Along the diagonal are the histograms of the estimated 1D marginal posterior probability distribution for each parameter. The recovered parameters are more accurate and precise  and the peaks of the histograms are much closer to the injected values for the UIE model (right panel), while the CIE template (left panel) does not recover the injected signal value very well.
}
\label{posteriormargin}
\end{figure*}

To show the effects of the templates on parameter inference, we simulate the data by injecting an UIE signal generated by Eqs.~(\ref{template}) and (\ref{eq:xnr}) into simulated Gaussian noises. For the injection, we choose $A = 1.5\times10^{-21}$, $\phi = 0$, $f \simeq 250 Hz$, and $\tau\simeq4\times10^{-3}s$, so that $\Psi^{BH}(t)$ corresponds to the ringdown signal of a 68 $M_\odot$ GR black hole. For the echo signals, we assume ${\cal A} = 0.5$, $\beta = 0.006$, $t_{echo} = 0.0295$, $r=0.15$, and $\Delta t_{echo} = 0.0295$. We have fixed the number of echoes to be $\tilde N_{echo}=10$ for the sake of simplicity. In principle, one can include more artificial echoes in the train, but the amplitude decreases quickly for later echoes. So 10 echoes are already sufficient to capture the dominating effects of the signal on the signal to noise ratio (SNR). In this case, the SNR is 18, greater than the threshold of making a detection. For the noise, we consider the forecast noise curve for Advanced LIGO at design sensitivity \cite{final} to simulate the noise-limited constraints that can be obtained by Advanced LIGO for a GW150914-like event. Then we fit the simulated data with both the CIE and UIE template. We sample the likelihood function using the {\it emcee} package \cite{ForemanMackey:2012ig} on the 4-dimensional and 5-dimensional parameter space for the CIE and UIE template respectively, with a prior shown in Tab.~\ref{tab:prior}. The marginalized $1\sigma$, $2\sigma$,  and $3\sigma$ constraints on the parameters are shown in Fig.~\ref{posteriormargin}. The parameters inferred from the MCMC sampling are shown in Tab.~\ref{summary}.

As we can see from Fig \ref{posteriormargin}, the UIE template can recover the injected signal very well with the maximum a posterior estimations (MAPs) very close to the injected values. For the CIE template, the MAPs are relatively far away from the values of the injected signal. Also as shown in Tab \ref{summary}, the standard errors and mean values of the CIE model are not very satisfactory. From this simulation, we see that it can lead to large errors to use the simple CIE template to exact echo signals for some cases. If there is a UIE signal in the real data, using a CIE template may incorrectly assign a low statistical significance for the UIE signal, falsely excluding the model. On the other hand, one also has to be careful when interpreting in the opposite way: a better posterior marginal contour plot is not sufficient to claim that the UIE template is favored comparing to the CIE template, as the UIE template has more degrees of freedom and therefore should be penalized in model selection.

\begin{table}[tbp]
\begin{tabular}{lcc}
\hline\hline
 ~ & $\bf Prior\; range$ & $\bf Injection\; value$\\
\hline
\ ${\cal  A}$  & ${(0,1)}$ & $0.5$\\
\ $\beta $  & ${(0.001,0.01)}$ & $0.006 $\\
\ $t_{echo}$           & ${(0.02,0.04)}$ & $0.0295 $\\
\ $r$ & ${(0.1,0.2)}$ & $0.15 $\\
\ $\Delta t_{echo}$            & ${(0.02,0.04)}$ & $0.0295 $\\
\hline\hline
\end{tabular}
\caption{The prior range of echo parameters. The prior distribution of each parameter is uniform over the respective prior range. There is an upper limit on $A$'s prior range since the amplitude of echoes can not be greater than the amplitude of the IMR signal. $\beta$ is the width of the echo Gaussian profile. $t_{echo}$ and $\Delta t_{echo}$ are approximately the same, both of which are allowed to vary independently within $1\%$ of their respective maximum values. $r$ is defined in Eq \ref{eq:xnr}.}
\label{tab:prior}
\end{table}

When the CIE template fails and a more general UIE template becomes necessary, the assessment can be done with the Bayesian model selection. Let us denote the CIE and UIE templates as hypothesis ${\cal H}_0$ and ${\cal H}_1$, and assume observed data $d(t)$. The logarithm of the Bayes factor $B_{01}$ is defined as
\be
\ln B_{01} = \ln p(d|{\cal H}_1,I) - \ln p(d|{\cal H}_0,I),
\ee
where $\ln p(d|{\cal H}_i,I)$ is known as the evidence for hypothesis ${\cal H}_i$, and can be calculated through Eq.~(\ref{likelihood}) with $\theta$ being integrated over the parameter space.

Practically, $\ln p(d|{\cal H}_i,I)$ is computed by numerically integrating over parameter space $\theta_i$ of hypothesis ${\cal H}_i$ using sampling algorithms such as Parallel-Tempering MCMC with thermodynamic integration; see \cite{chi:2013chi} for more details. Again, we use the forecast noise curve for Advanced LIGO at design sensitivity.

\begin{table}[tbp]
\begin{tabular}{lcc}
\hline\hline
Parameters & CIE &UIE \\
\hline
${\cal A}$  &$0.3625^{+0.167}_{-0.099}$&$0.500^{+0.138}_{-0.120}$\\
$\beta$  &$0.0061^{+0.0028}_{-0.0021}$ &$0.0060^{+0.0020}_{-0.0016}$\\
$t_{echo}$ & $0.037^{+0.0037}_{-0.0166}$&$0.0295^{+0.0003}_{-0.0003}$\\
$r$    & $--$&$0.150^{+0.003}_{-0.003}$   \\
$\Delta t_{echo}$  &$0.0296^{+0.0006}_{-0.0002}$&$0.0295^{+0.0002}_{-0.0002}$ \\
\hline\hline
\end{tabular}
\caption{The summary of the statistics of a parameter estimation run with a $3\sigma$ error bar.}
\label{summary}
\end{table}

\begin{figure}[tbp]
\includegraphics[width=0.5\textwidth]{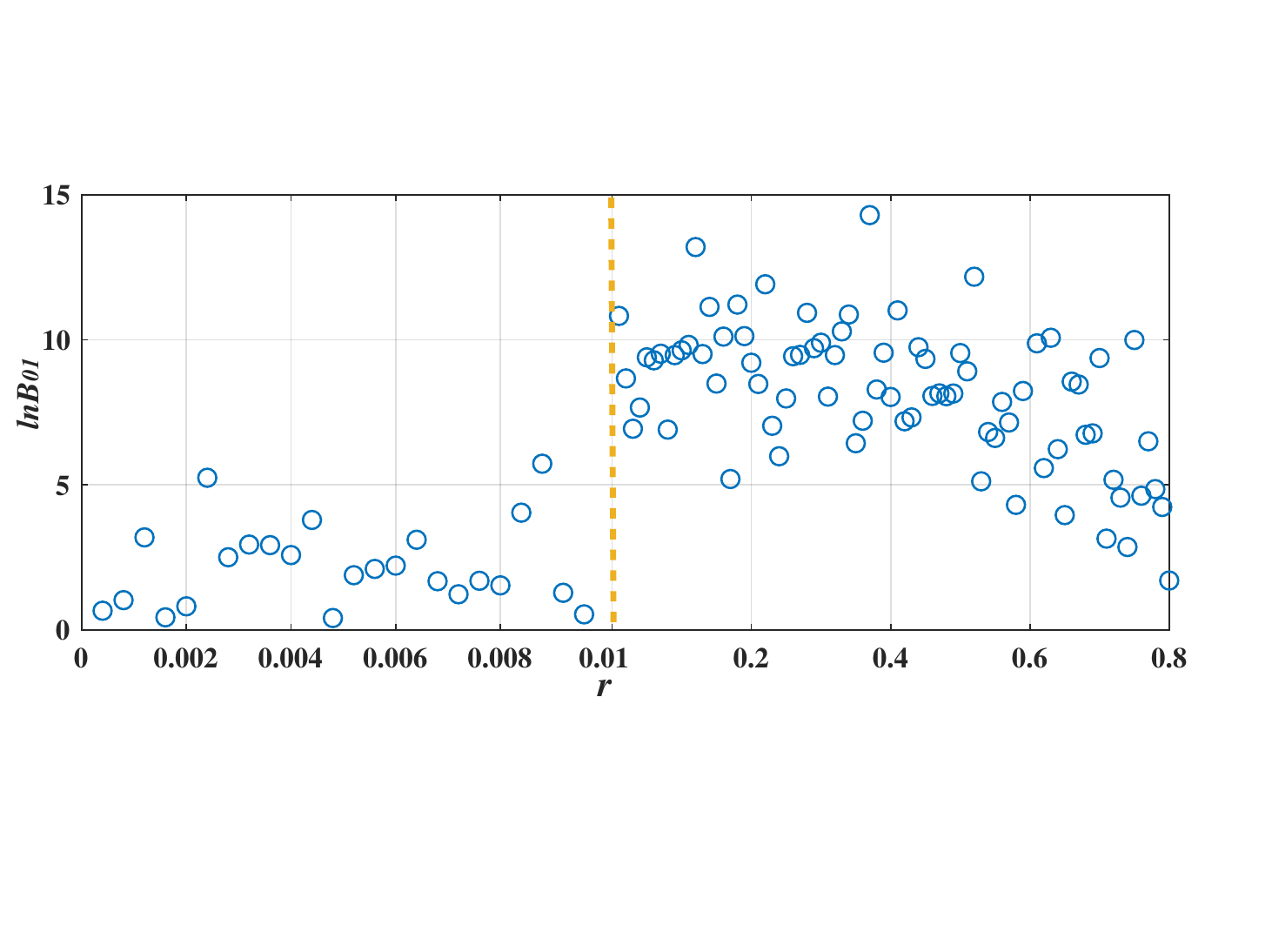}
\includegraphics[width=0.45\textwidth]{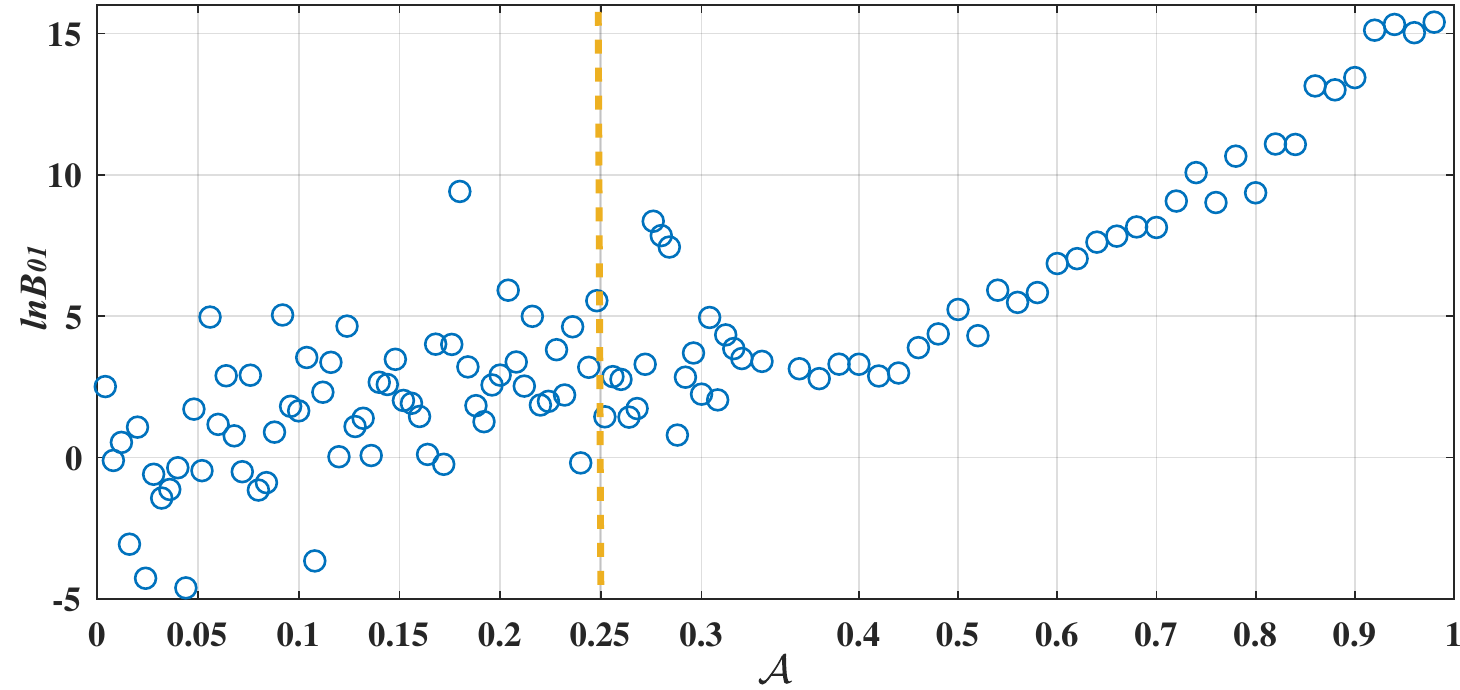}
\caption{The two panels are distribution{\red s} of $\ln B_{01}$ for different $r$ (with ${\cal A} = 0.6$) and ${\cal A}$ (with $r=0.15$) respectively. The other parameters are chosen as $\beta = 0.006$, $t_{echo}=0.0295$, and $\Delta t_{echo} = 0.0295$. The yellow dashed lines in both panels are the thresholds of $r$ and ${\cal A}$ where the Bayes factor can be used to select the UIE model.}
\label{scatterfrequency}
\end{figure}

The results are shown in Fig.~\ref{scatterfrequency}. The upper panel of Fig.~\ref{scatterfrequency} shows $\ln B_{01}$ as a function of $r$ with other echo parameters fixed to ${\cal A} = 0.5$, $\beta = 0.006$, $t_{echo}=0.0295$, and $\Delta t_{echo} = 0.0295$. We can see that $\ln B_{01}$ is all above $0$, which means that the UIE model is generally favored in the Bayesian model selection by at least $2\sigma$ in statistical significance. As one may expect, for small $r$, the significance is not that obvious. The significance increases as $r$ increases. Particularly, the UIE template is favored by $3\sigma$ when $r > 0.1$, which indicates that the UIE template is necessary in this case. We can also find that the significance starts decrease at large $r$, which seems to be due to that the increasing of the echo interval leads to a dramatic decrease in the echo amplitude. As a result, signals with large $r$ have relatively low SNR and hence low significance. The bottom panel of Fig.~\ref{scatterfrequency} shows $\ln B_{01}$ as a function of ${\cal A}$, which closely relates to the SNR. In this plot, the other echo parameters are fixed to $\beta = 0.006$, $t_{echo}=0.0295$, $r=0.15$, and $\Delta t_{echo} = 0.0295$. We can see that, for small ${\cal A}$, the signal is not loud enough to distinguish UIE from CIE. The significance decreases as ${\cal A}$ decreases, and when ${\cal A} < 0.25$, which corresponds to a SNR of $8.76$, we find there will be some values of ${\cal A}$ that $\ln B_{01} < 0$, so we can not distinguish the CIE and UIE model in this region. Excluding this region, however, one can distinguish the two templates at $3\sigma$. $\ln B_{01}$ being sensitive to the value of ${\cal A}$ also supports our explanation on the decreasing of $\ln B_{01}$ at large $r$ shown in panel of Fig.~\ref{scatterfrequency}.

\section{Forecast on parameter constraints}\label{Sec:Fisher_matrix}

Have shown the necessity of the UIE template in a generic echo data analysis, in this section we discuss the constraint{\red s} on the parameters, especially on $r$, given by future GW detectors. The constraints on the parameters can be estimated with the Fisher information matrix. The Fisher information matrix $\Gamma_{ij}$, which characterizes the curvature of the likelihood function, can be defined in terms of the partial derivatives of the GW template with respect to the echo parameters,
\begin{equation}
\Gamma_{ij}=4\int^{{f_{\rm max}}}_{{f_{\rm min}}}df\frac{1}{S_{n}(f)}\frac{\partial\tilde {h}^{*}(f)}{\partial\theta^i}\frac{\partial \tilde{h}(f)}{\partial\theta^j}.
\end{equation}
The statistical error of the parameter $\theta^{i}$ can be estimated as
\ba
\sigma_i = \sqrt{\Sigma_{ii}}
\ea
where $\Sigma_{ii} = (\Gamma_{ii})^{-1}$ is the diagonal element of the inverse of the Fisher matrix. For the noise curve, we utilize the sensitivity of future generations of detectors, Advanced LIGO with anticipated final design sensitivity \cite{final} as well as Einstein Telescope (ET) \cite{Hild:2010id}.

For the injection we considered above, the constraints estimated using the Fisher matrix approach are shown in Tab.{\ref{fisher}}, which agrees with the result we obtained from the MCMC sampling.

\begin{table}[tbp]
\begin{tabular}{lc}
\hline\hline
Parameters & result \\
\hline
${\cal A}$  &$0.5\pm0.032$\\
$\beta$  &$0.006\pm4.6\times10^{-4}$ \\
$t_{echo}$ & $0.0295\pm2.95\times10^{-7}$\\
$r$    & $0.15\pm1.5\times10^{-4}$  \\
$\Delta t_{echo}$  &$0.0295\pm8.85\times10^{-6}$\\
\hline\hline
\end{tabular}
\caption{The summary of UIE parameter estimation with fisher information matrix. We use the simulation noise generate from Advanced LIGO \cite{final}, and the injected value can be found in Tab.{\ref{tab:prior}}.}
\label{fisher}
\end{table}
Moreover, we can discuss the dependence of constraints on the physical events. We will mainly focus on constraints on $\Delta\,t_{echo}$ in the CIE template and $r$ in the UIE template, and will discuss the dependence on other parameters. The results are presented in terms of the relative error $\epsilon_{i}=\sigma_{i}/\theta^i$.

\begin{figure}[tbp]
\includegraphics[height=0.165\textwidth]{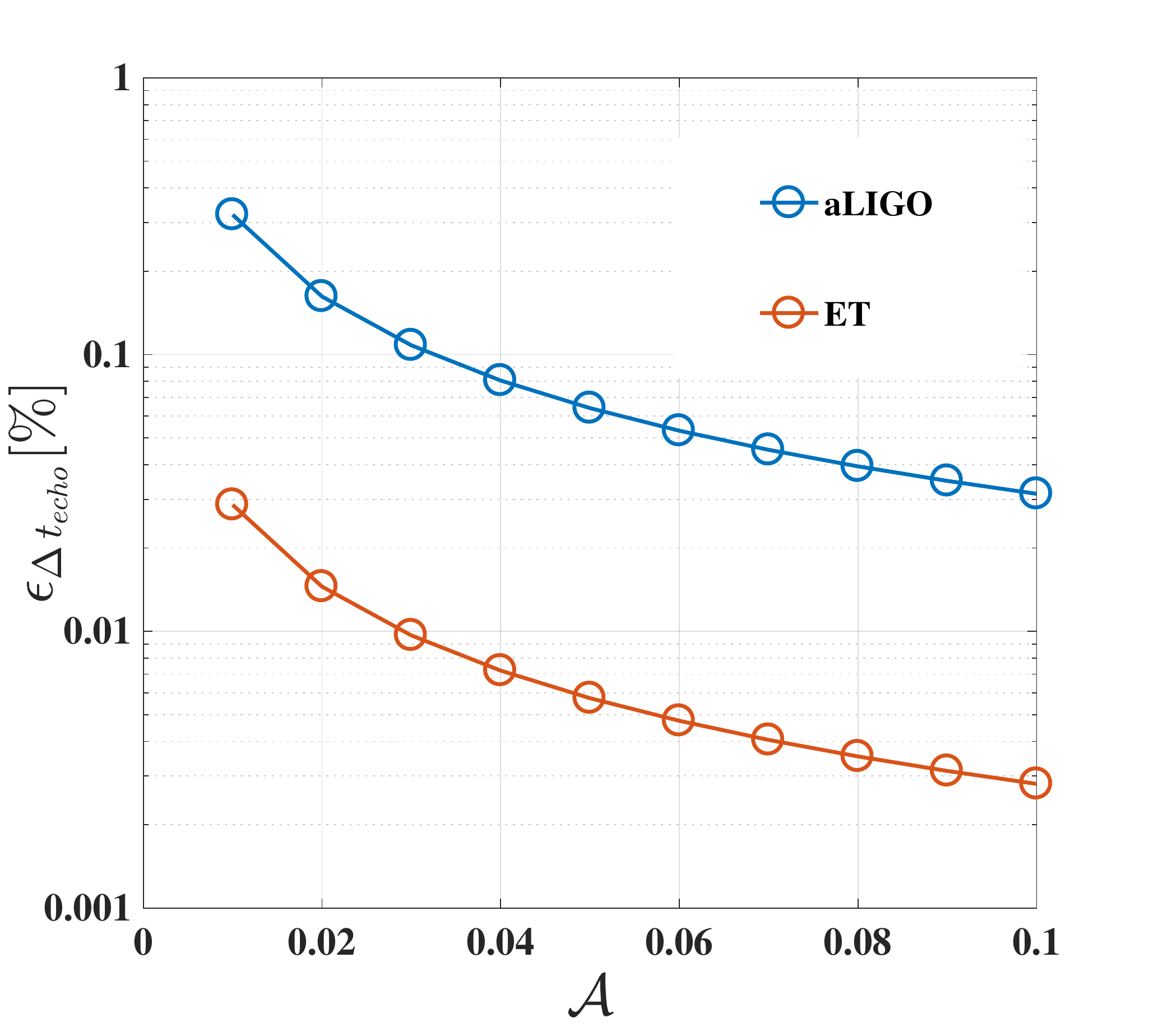}
\includegraphics[height=0.165\textwidth]{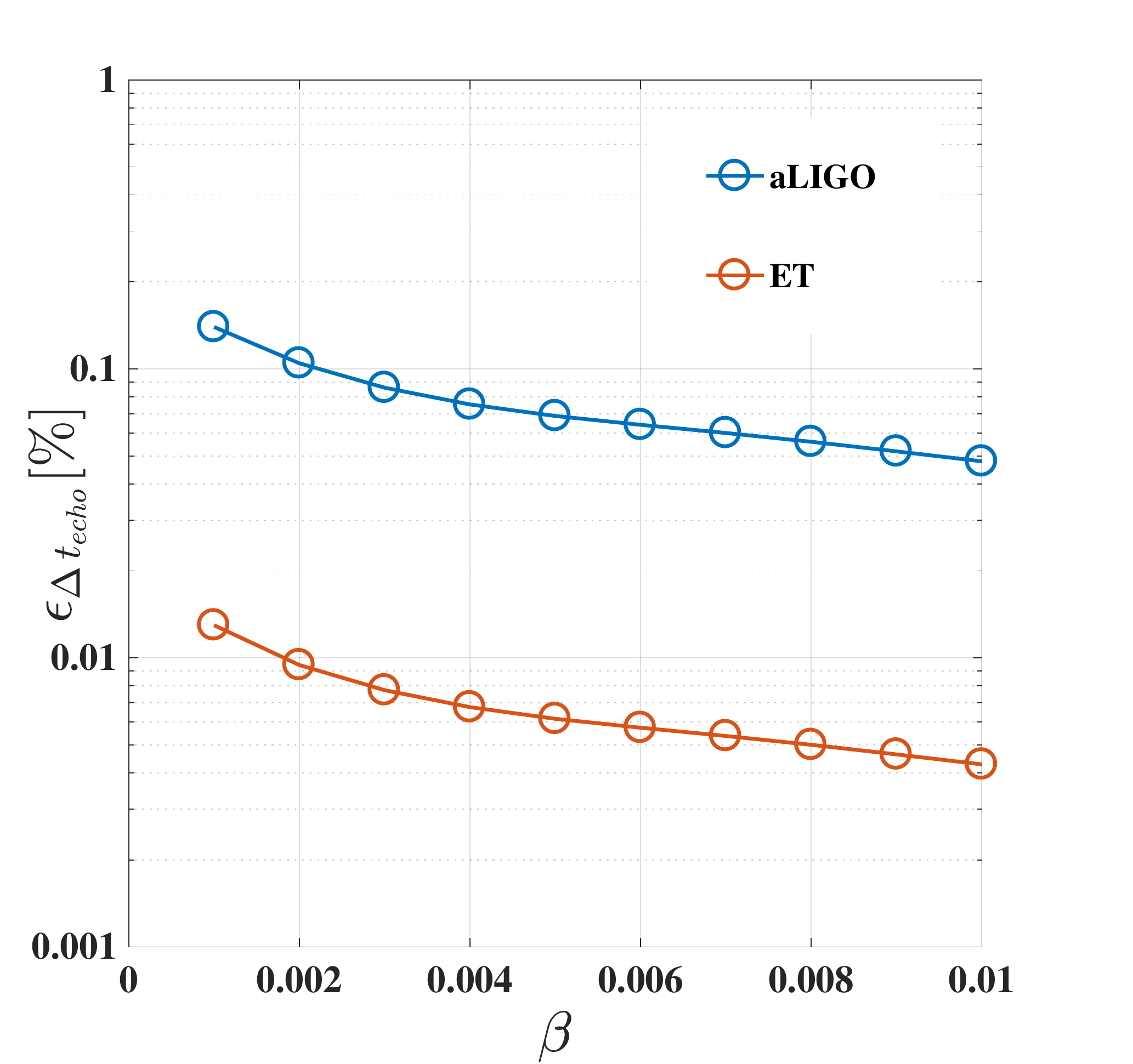}
\includegraphics[height=0.165\textwidth]{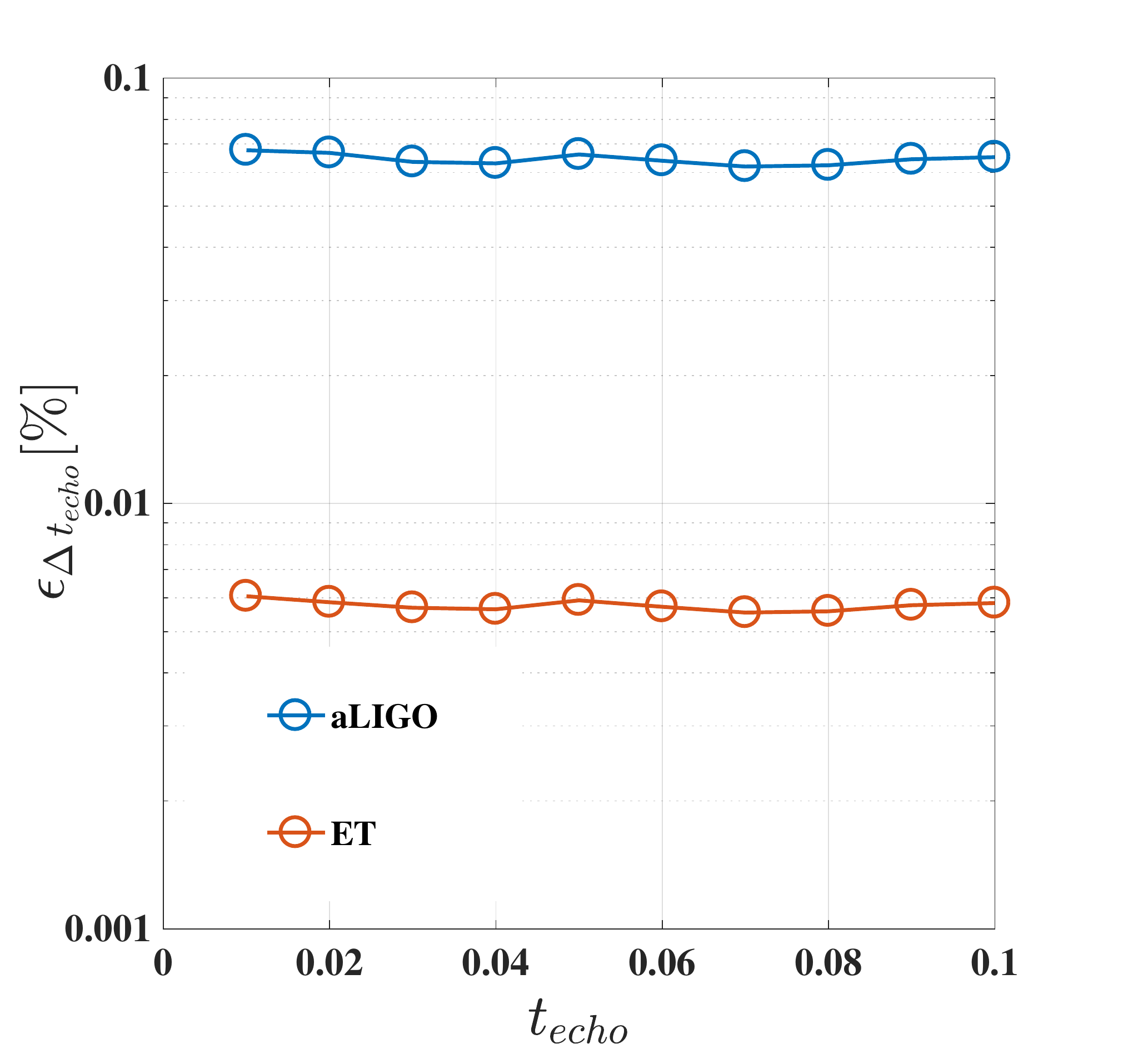}
\includegraphics[height=0.165\textwidth]{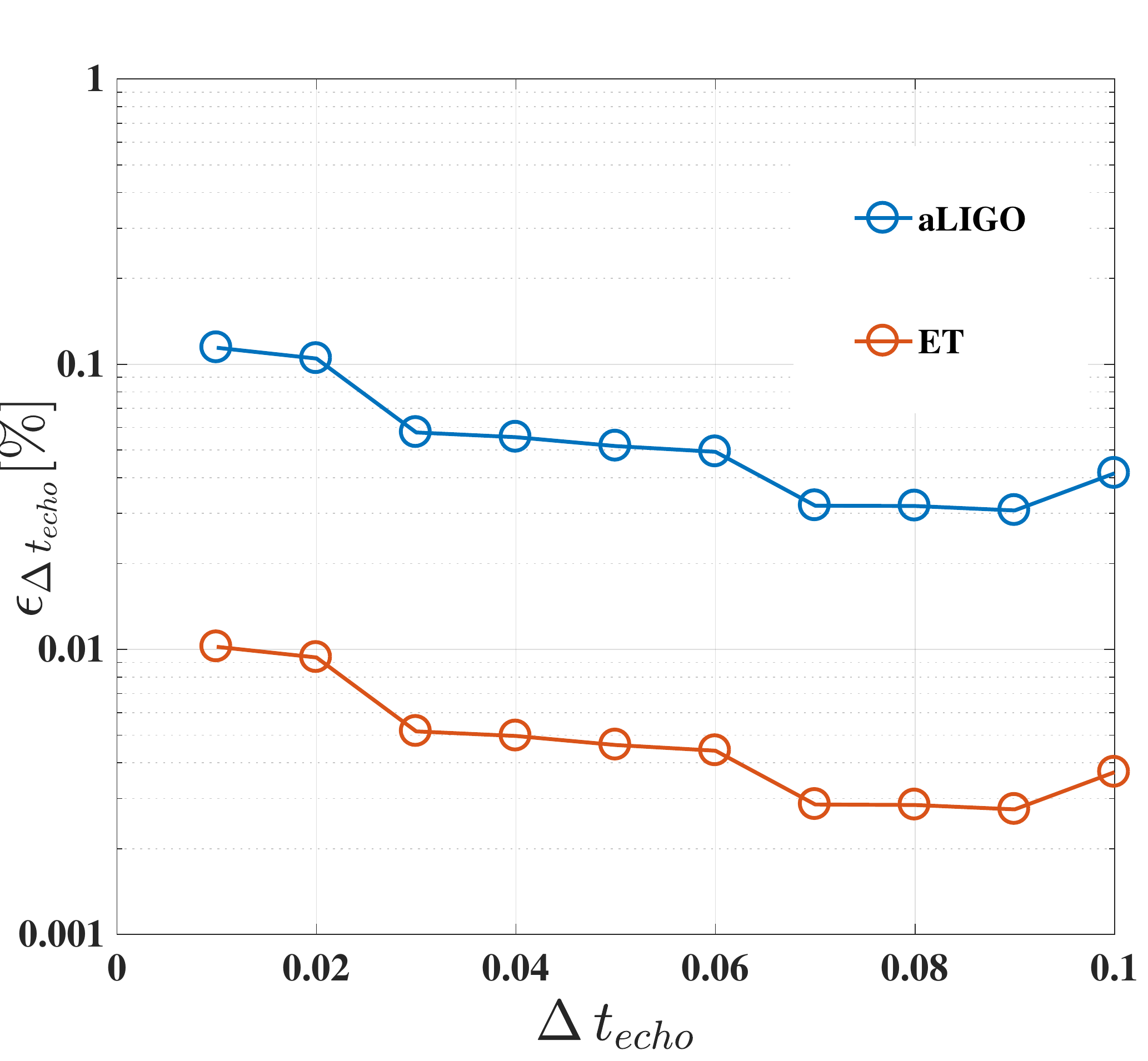}
\caption{Relative error on $\Delta\,t_{echo}$ for CIE templates, computed with forecast Advanced LIGO at final design sensitivity (blue) and ET with the forecast sensitivity (red).}
\label{CIEfisher}
\end{figure}

\begin{figure}[tbp]
\includegraphics[height=0.165\textwidth]{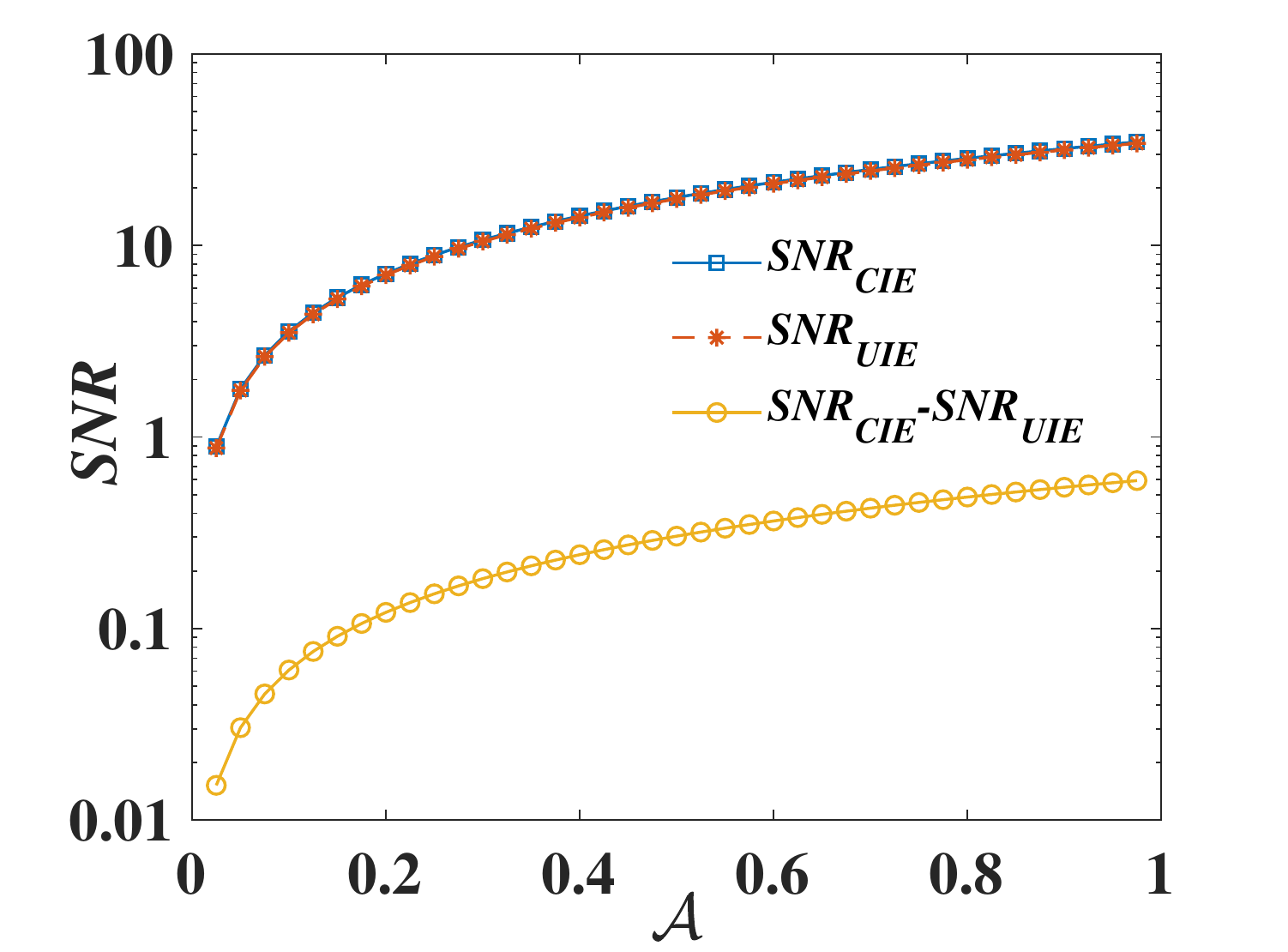}
\includegraphics[height=0.165\textwidth]{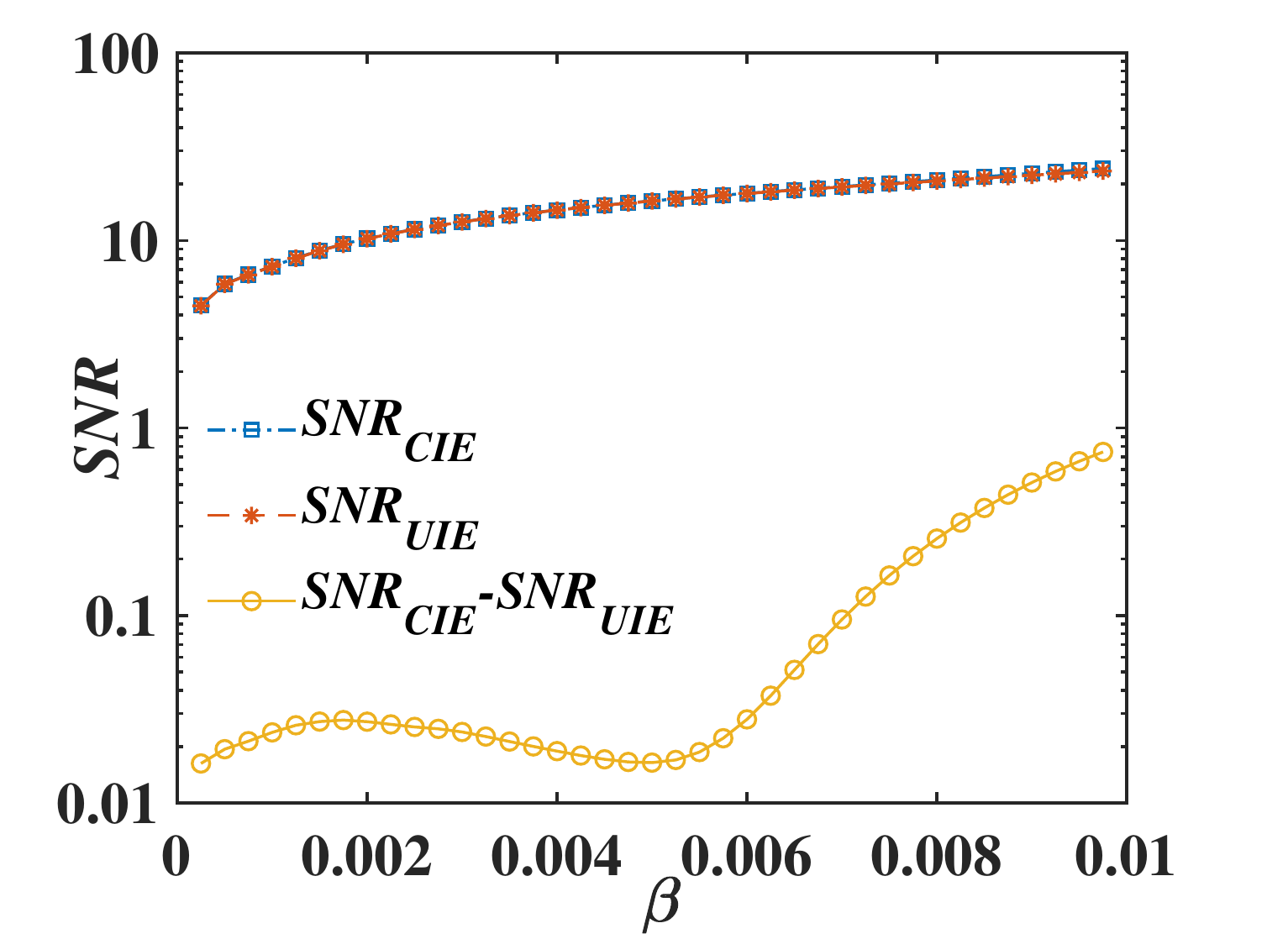}

\caption{The Signal to Noise Ratio of the CIE and UIE model, in the left panel we fixed $\beta = 0.006$s, $\Delta t_{echo}= 0.0295$s, $r=0.15$, and $t_{echo} = 0.0295$s and in the right panel we fixed ${\cal A} = 0.5$,  $\Delta t_{echo}= 0.0295$s, $r=0.15$, and $t_{echo} = 0.0295$s .}
\label{SNR}
\end{figure}

Fig.~\ref{CIEfisher} shows the dependence of $\epsilon_{\Delta\,t_{echo}}$ on the parameters of the CIE template. In each plot, we fix the parameters to be the injection values shown in Tab.~\ref{tab:prior}, except for the parameter labeled in the horizontal axis. The relative error of $\Delta\,t_{echo}$ decreases as the value of the relevant parameter increases. We can see that the detection precision of $\Delta\,t_{echo}$ increases significantly as ${\cal A}$ or $\beta$ increase. It is because the SNR of the signal is enhanced by large ${\cal A}$ or $\beta$. As shown in Fig.~\ref{SNR}, for ${\cal A} = 0.1$, we have SNR $= 3.56$, and for ${\cal A} = 1$, we have SNR $= 34.7$. Similarly, the SNR corresponding to $\beta = 0.001$ is 7.28, which increases to $24.68$ when $\beta =0.01$. On the other hand, $\epsilon_{\Delta\,t_{echo}}$ is not very sensitive to the other two parameters.


\begin{figure}[tbp]
\includegraphics[height=0.165\textwidth]{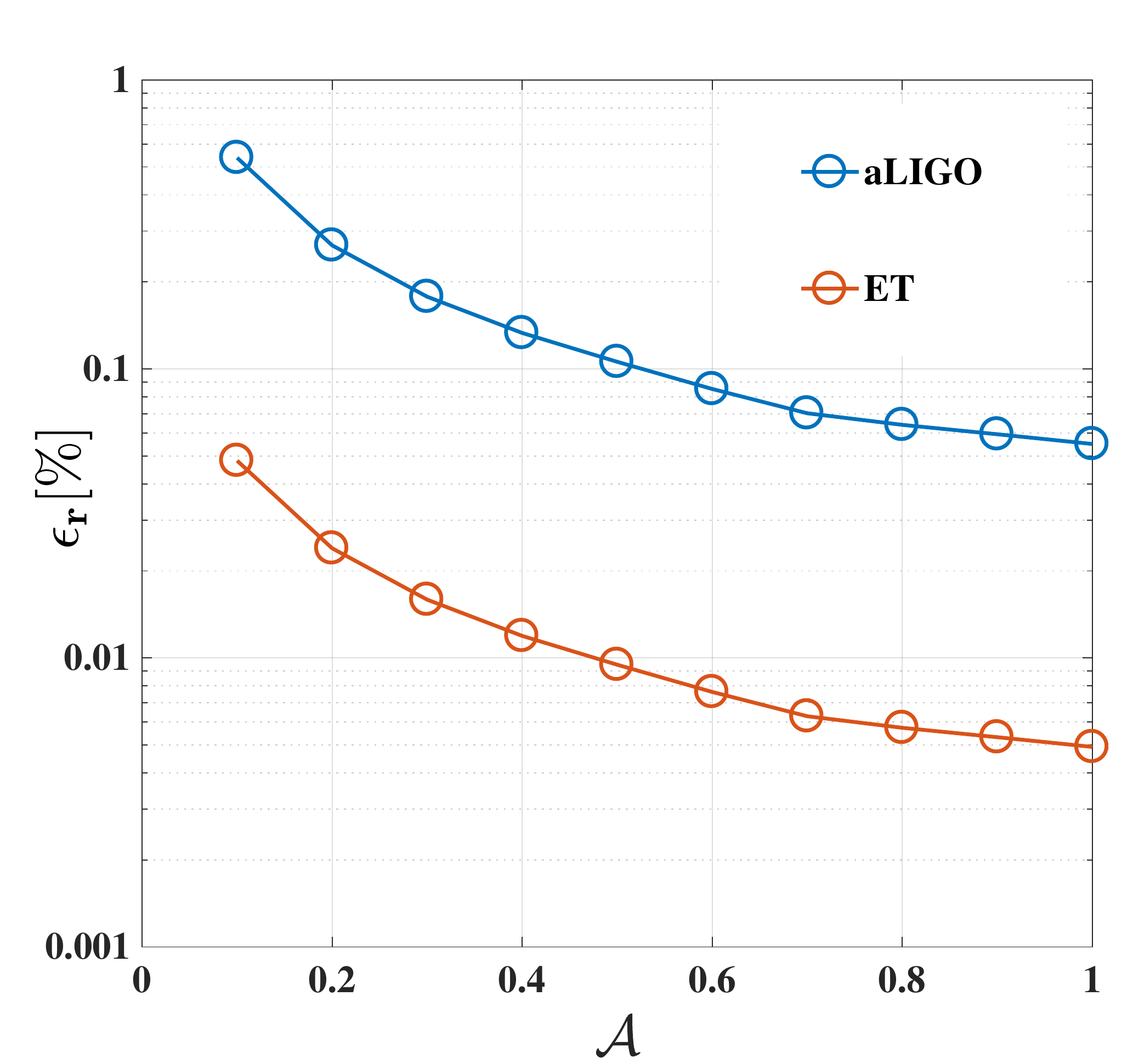}
\includegraphics[height=0.165\textwidth]{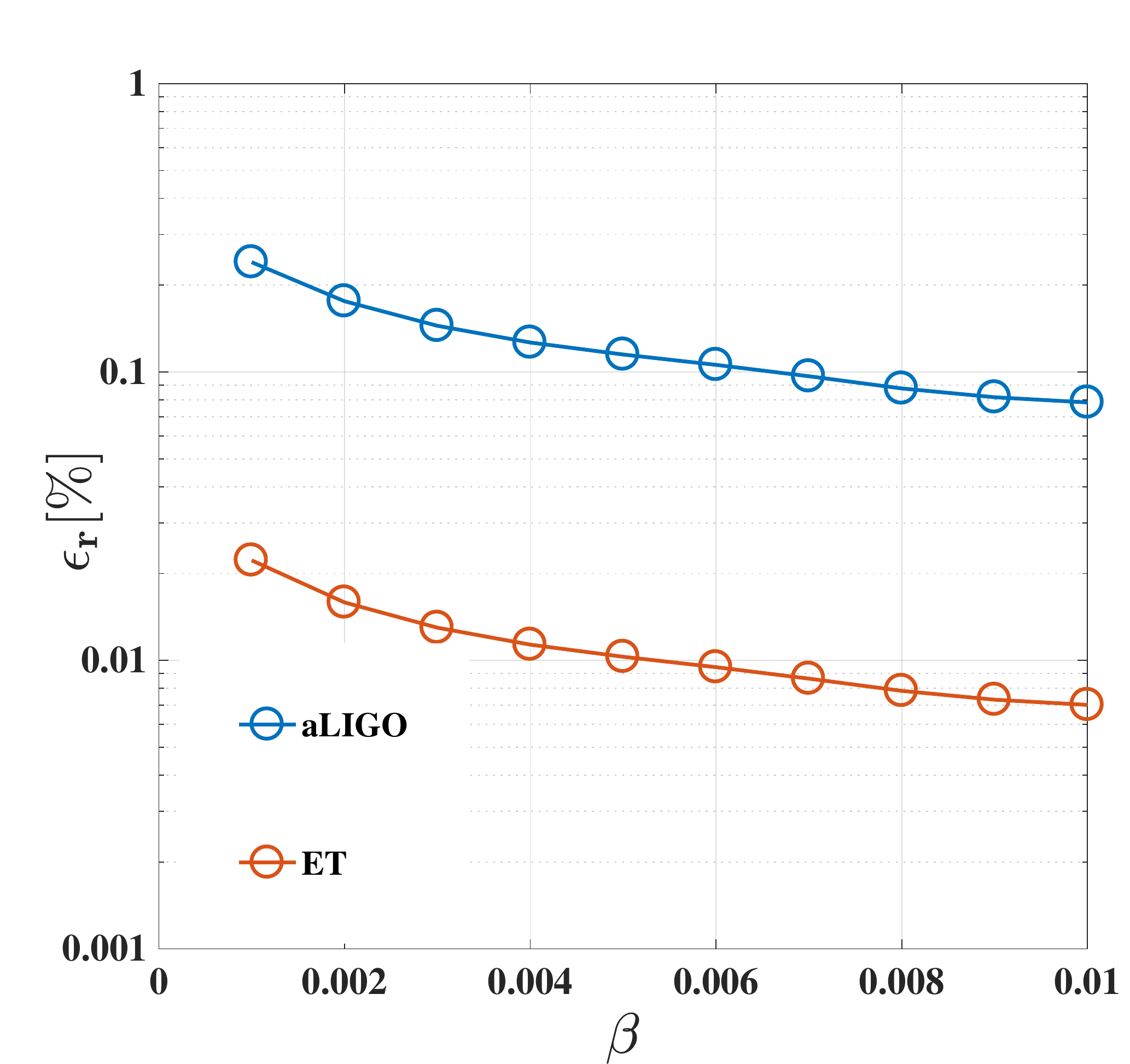}
\includegraphics[height=0.165\textwidth]{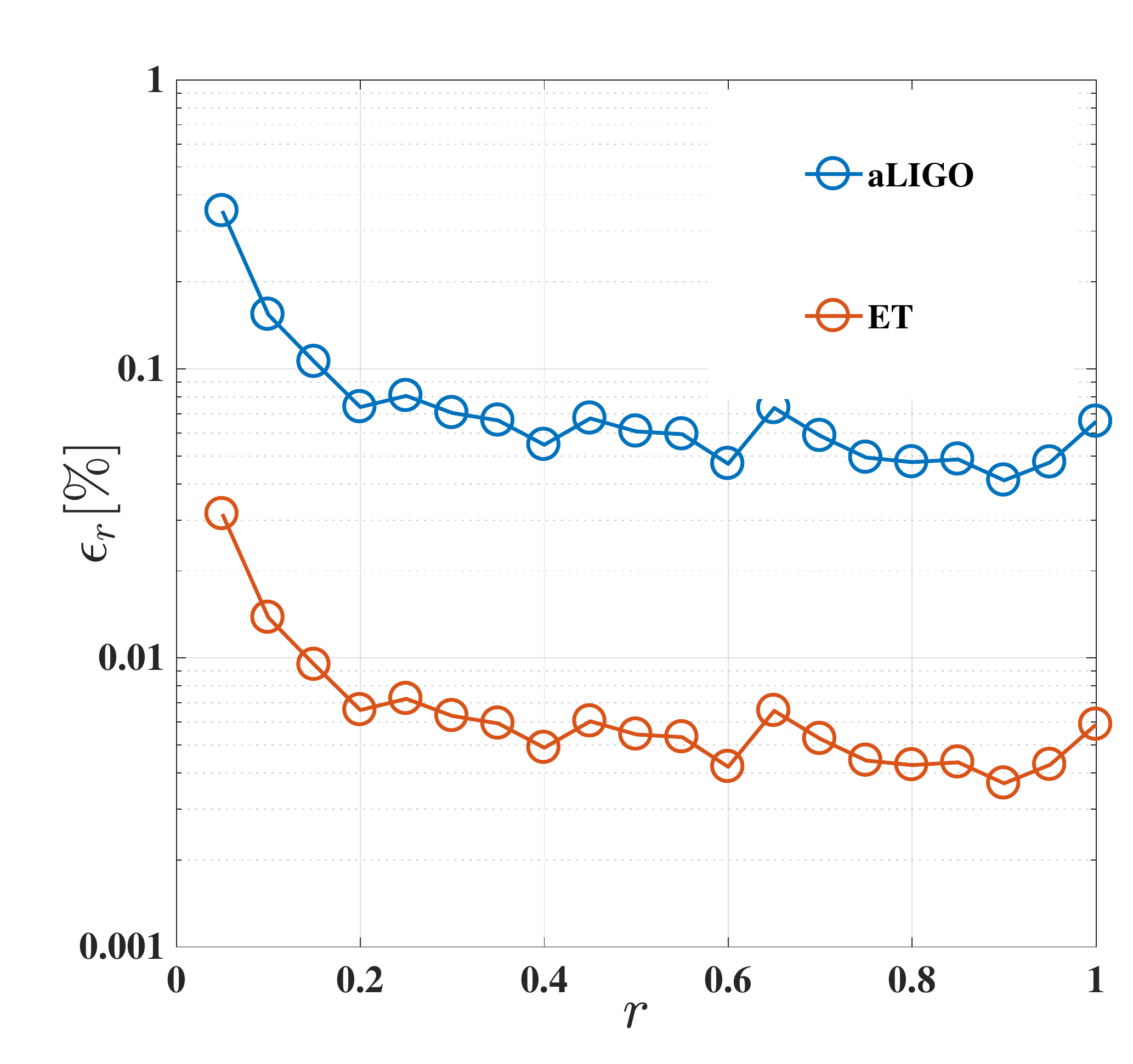}
\includegraphics[height=0.165\textwidth]{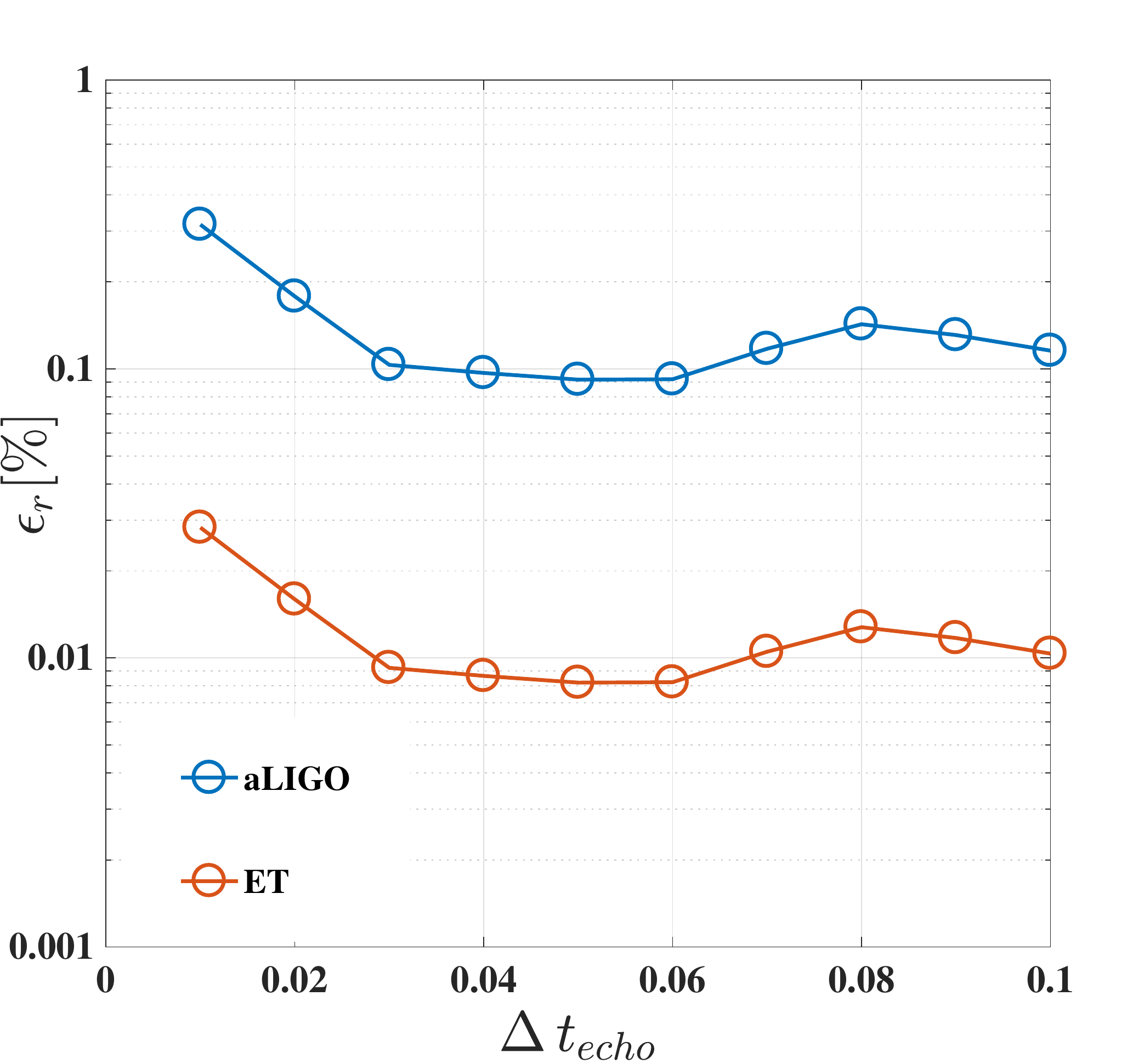}
\caption{Relative errors on $r$ in the UIE template, computed with forecast Advanced LIGO at final design sensitivity (blue) and ET with the forecast sensitivity (red). }
\label{UIEfisher}
\end{figure}

Fig.~\ref{UIEfisher} shows the dependence of $\epsilon_r$ on the parameters in UIE templates.
We see that the detection precision of $r$ increases as the SNR of the signal increases with $\cal A$ and $\beta$. Moreover, the detection precision of $r$ increases significantly as $r$ cross $0.1$ from below. The reason could be that, as we discussed in Sec.~\ref{Sec:Parameter Inference}, the UIE template becomes distinguishable from the CIE template for $r>0.1$.

Given these result, we may expect that the Advanced LIGO (with design sensitive) and future GW detectors can not only detect or exclude GW echoes, but also can fruitfully extract information from the echo signals if they are detected. Especially, with the next generation GW detector, such as ET, the relative error on $\Delta\,t_{echo}$ will improve by more than one order of magnitude. As we discussed in Sec. \ref{Sec:Parameter Inference}, the UIE template become necessary when $r > \epsilon_{\Delta\, t_{echo}}$, our results show that a well modelled echo interval template is very important in the echo searching with the next generation GW detectors.

\section{Discussions}\label{Sec:Discussions}

In this paper, we {have investigated} whether and when the UIE template is needed in GW echo searches. It is important as the using of CIE templates needs to be properly justified in the absence of any concrete ECO model, in which case the exact echo waveform is unknown. Especially, there are mechanisms that can generate echoes with unequal intervals \cite{Wang:2018mlp,Wang:2018cum}.

We first used the MCMC sampling to reconstruct injected UIE signals in Gaussian noises with both UIE and CIE templates, showing that the CIE template may mis-reconstruct the signal. We further supported this result by performing the Bayesian analysis on model selection. In particular, we have studied the dependence of the Bayes factor on the echo signals. In terms of the interval change ratio $r$, we found a window in which the UIE template can be significantly distinguished from the CIE template. For parameter space outside this window, the two template cannot be well distinguished either because the difference between the two template is negligible (small $r$) or the echo signal is not loud enough (large $r$). We found that a GW detector like Advanced LIGO (at design sensitivity) can distinguish two templates at $3\sigma$ given a GW150914-like event with $r > 0.01$ and other parameters fixed as in Fig.~\ref{scatterfrequency}. The statistical significance on distinguishing these two templates is sensitive to the amplitude of the echo as well. Given a similar event we consider before, we found that for $r = 0.15$, the Advanced LIGO (at design sensitivity) can distinguish these two templates at $3\sigma$ if the first echo amplitude is larger than $25\%$ of that of the ringdown signal with other parameters fixed as in Fig.~\ref{scatterfrequency}, corresponding to a SNR of $8.76$.

Have shown the necessity of the UIE template, we have forecast the constraints on the echo intervals given by future GW detectors such as Advanced LIGO and ET. We have estimated the errors on the parameters of the UIE templates and the their dependence on the properties of the physical signals using the Fisher information matrix. We found that the full sensitivity Advanced LIGO (ET) can narrow the error on $\Delta\, t_{echo}$ down to $1\%$ ($0.1\%$) or even tighter. In this case, a small change in the echo interval can have considerable effects on parameter inference. Our study suggests that echo search template bank should be enlarged with UIE templates so that a proper inference can be achieved when more general phenomena of ECOs are considered.

\textbf{Acknowledgments}

YTW is supported in part This project is supported by National Natural Science Foundation of China(Grant No.11805207), by the sixty-second batch of China Postdoctoral Fund. JZ is supported by European Union's Horizon 2020 Research Council grant 724659 MassiveCosmo ERC-2016-COG. SYZ acknowledges support from the starting grant from University of Science and Technology of China (KY2030000089) and the National 1000 Young Talents Program of China (GG2030040375). YSP is supported by NSFC, Nos. 11575188, 11690021,and also supported by the Strategic Priority Research Program of CAS,No. XDB23010100.

 \end{document}